# First-Principles Insights into Metallic Doping Effects on Yttrium $\{10\bar{1}0\}$ Grain Boundary

Guanlin lyu [a], Yuguo Sun [a], Ping Qian [a,*], Panpan Gao [a,*]

[a] University of Science and Technology Beijing, Beijing 100083

**Abstract**

   Metallic yttrium and its alloys are promising materials for high-tech applications, particularly in aerospace and nuclear reactors. The doping of metallic elements at grain boundaries can significantly influence the stability, strength, and mechanical properties of these materials; however, studies on solute segregation effects in Y-based alloys remain scarce. To address this gap, this work employs first-principles calculations to systematically examine the effects of doping with 34 metallic elements on the properties of a highly symmetric twin grain boundary in metallic yttrium. All solute elements exhibit a tendency to segregate to regions near the grain boundary, driven by segregation energy. Additionally, energy barriers influence these elements to prefer segregation positions farther from the grain boundary line. the strengthening energy calculations reveal that all dopant elements enhance grain boundary strength when located near the boundary. For grain boundary energy and solubility trends, elements within the same transition metal group across different periods display consistent behaviors. Considering segregation, strengthening, and grain boundary energy effects, we identify 11 elements (Al, Zn, Rh, Pd, Ag, Cd, Sn, Ir, Pt, Au, Hg) that preferentially segregate near the grain boundary, where they contribute to grain boundary strengthening and enhanced stability. By decomposing the strengthening energy into mechanical, chemical, and vacancy formation components, we find that the chemical contribution is the primary factor in strengthening, while the mechanical contribution of transition metals correlates with changes in the Voronoi volume and relative atomic radius of the solute. The density of states analysis indicates that increased grain boundary stability arises mainly from hybridization between solute d orbitals and yttrium, leading to more stable electronic states. This study provides theoretical guidance for optimizing metallic dopants in Y-based alloys.

**Keywords:** Segregation Engineering; Grain Boundary Segregation; Yttrium-Based Alloys; First-Principles Calculations; Grain Boundary Energy; Grain Boundary Strengthening

**Introduction**

Grain boundary(GB) is defect regions commonly found in alloy materials, directly influencing their tensile strength, fracture toughness, stability, and other mechanical properties[1-3]. Due to their complex atomic structures and high energy, GBs are a challenging and popular topic in materials research. Numerous studies indicate that segregation behavior at GBs significantly impacts the mechanical properties of alloys. Even trace amounts of dopants can have substantial effects, a phenomenon often referred to as "segregation engineering"[4,5]. In alloys, element segregation from the matrix to the GB can lead to intergranular embrittlement, a critical issue that warrants attention. Thus, further research is essential to expand material databases and enable the targeted selection of optimal dopants to modify properties such as material strength, ultimately guiding material design.

Atom probe tomography (APT) is a commonly used technique for studying GB segregation. Raabe et al.[6] employed APT to investigate the segregation and transformation mechanisms from nano-martensite to austenite, observing that Mn segregation at GBs facilitated elastic stress relief in the martensitic matrix and promoted phase transformation at martensitic GBs. Rosa et al.[7] used secondary ion mass spectrometry and APT to detect boron segregation at austenitic GBs in high-strength low-carbon steel, finding that boron segregation at γ-GB increased with temperature. While techniques such as APT and high-resolution transmission electron microscopy (HRTEM) provide critical insights into GB segregation, they are often limited by resolution and sample preparation constraints, hindering comprehensive analysis of complex atomic interactions and energy states. By contrast, first-principles calculations allow precise predictions of atomic-scale material structures, energies, and interactions.

Recently, many researchers have used first-principles methods to simulate the effects of dopants on GB properties in various alloy systems, such as Al[8-13], Ni[14-18],

Fe[19-22], Mg[23-24], W[25], and Cu[26-27]. Wu et al.[25] studied metallic dopants in tungsten GBs and found a correlation between dopant atom radius and segregation tendency, with dopant-induced bonding strength enhancement observed in high-energy GB systems. Huang et al.[26] investigated the segregation of different metallic dopants in the Cu Σ5(310) GB, revealing a link between the excess free volume at the GB and its energy. Large solute atoms or interstitial impurities filled the excess free volume, effectively reducing GB energy. Millett et al.[27] conducted a comprehensive survey on the segregation and GB energy effects of various metallic dopants in copper. Ito et al.[28] observed that manganese reduces cleavage fracture energy in iron GBs, with ferromagnetic coupling further enhancing this effect, emphasizing the role of local magnetic states in manganese-containing steels. Other studies have identified correlations between dopant segregation behavior and atomic size, electronic structure, and free volume changes. For example, Hu et al.[21] analyzed solute effects on iron GBs, finding that segregation energy depends on both strain and electronic factors. Pei et al.[23] studied solute segregation at Mg $\{10\bar{1}1\}$ and $\{10\bar{1}2\}$ GBs, noting that solute volume significantly influences segregation energy. Liang et al.[29] systematically investigated the segregation behavior of O and H impurity atoms at the $\{10\bar{1}2\}$, $\{11\bar{2}1\}$, $\{11\bar{2}2\}$, and $\{10\bar{1}1\}$ grain boundaries in Ti using first-principles calculations. They observed a strong segregation tendency for H impurities at all four Ti grain boundaries, whereas O impurities showed no significant segregation at the $\{11\bar{2}2\}$ grain boundary but did segregate at the other three boundaries. Zheng et al.[30] constructed a dataset predicting the GB energy of hcp-Y Σ7(0001) twist GBs at 0K as 0.22 J/m² using first-principles calculations across 327 GBs of 58 metals. Hong et al.[24] explored the segregation effects of 19 solutes at Mg twin GBs, finding that GB segregation reduced lattice distortion and strengthened bonds near twin boundaries. Wang et al.[22] studied the Σ5(310) GB in bcc-iron and observed a negative linear relationship between solute segregation energy and atomic radius, showing that atomic size and electronic structure are crucial for GB strength enhancement. Despite extensive studies on dopant effects in various systems, research on metallic element influence on hcp-Y GB properties remains limited. This gap is particularly relevant for new energy applications, where yttrium alloys play crucial roles in aerospace, nuclear, and superconducting materials. Thus, advancing the understanding and design of Y-based alloys is of great significance.

In this study, first-principles calculations are used to systematically analyze the effects of metallic dopants on segregation at the GB of hcp-Y, aiming to guide the selection of optimal dopants in Y-based alloy design. We focus on the $\{10\bar{1}0\}$ twin GB in hcp-Y, which exhibits high strength, toughness, and favorable diffusion characteristics, and evaluate the influence of solute segregation on GB stability and strength by doping with 34 metallic elements. This study elucidates the mechanisms by which these elements impact GB stability, providing valuable insights for alloy processing.

**Computational Methods and Details**

In this study, first-principles calculations based on density functional theory (DFT) are conducted to investigate the atomic-scale properties of the $\{10\bar{1}0\}$ twin-grain boundary in the rare earth metal yttrium (Y). The computational approach employs well-established software packages widely utilized in scientific research, including the Vienna Ab initio Simulation Package (VASP 5.4.4)[31-33], ASE[34], and VESTA[35]. For the VASP calculations, pseudopotentials were selected from the recommended POTCAR files available on the VASP official website. The exchange-correlation interactions were treated using the generalized gradient approximation (GGA) with the Perdew-Burke-Ernzerhof (PBE) functional[36], and the projector augmented wave (PAW) method was employed to describe the ion-electron interactions[37]. A plane-wave cutoff energy of 400 eV is applied, with an energy convergence criterion (EDIFF) set to $1\times10^{-6}$ eV. During structural relaxation, the force convergence criterion ensures that the maximum force on each atom does not exceed 0.01eV/Å (EDIFFG = -0.01). The k-point mesh is generated automatically using a Monkhorst-Pack grid centered at the Gamma point, with a k-spacing value of 0.2Å$^{-1}$. For the free surface model calculations, the k-point mesh is consistent with that of the grain boundary, using a 9×3×1 grid. Spin polarization is enabled in calculations involving Fe, Co, Ni, Cr, and Mn, as these elements exhibit magnetic properties, to accurately determine their ground-state properties. At 0 K, the ground-state structure of metallic yttrium (Y) is hexagonal close-packed (hcp). Following full structural optimization, the lattice parameters of the Y unit cell are reported in Table 1. These parameters show good agreement with Spedding's research[47], with a volume deviation of approximately 1.1% and a lattice constant deviation of less than 5%. Based on these optimized parameters, bulk, grain boundary, and surface models for metallic Y are constructed. The bulk Y model is constructed by expanding the hcp unit cell along the three axes by a factor of 4 × 3 × 3, resulting in a model containing 72 atoms.

Fig.1 shows a three-dimensional orthogonal model of the twin grain boundary $\{10\bar{1}0\}$ in metallic yttrium, with the crystal directions along the x-axis, y-axis, and z-axis denoted as $<\bar{1}2\bar{1}0>$, $<0001>$, and $<10\bar{1}0>$, respectively. The lattice constants are 3.6Å × 11.2Å × 46.4Å, with 23 atomic layers along the z-axis. A total of 46 atoms are present, with 11 atomic layers above and below the grain boundary. To prevent interactions between adjacent grains due to periodic boundary conditions and to simulate the bulk metallic environment, adjustments were made to the grain boundary model. Specifically, a 6 Å vacuum layer was added to both the top and bottom of the model. Additionally, during lattice relaxation, the top and bottom 4 atomic layers along the z-axis were fixed[26,38], with the fixed atoms indicated in gray. Solute atoms, placed at substitutional positions, are marked in purple. The two surface models on the right side of Fig. 1 represent the upper and lower surfaces formed by the fracture of the twin grain boundary along the boundary line, corresponding to the $[10\bar{1}0]$ surface of yttrium. A 6Å vacuum layer was also applied to both the top and bottom of the surface models, with fixed atoms shown in gray. During structural optimization using first-principles methods, the atomic positions and lattice constants of all atoms, except those fixed to simulate the bulk environment, were optimized in the metallic Y unit cell model, the 4×3×3 supercell, the grain boundary model, the surface model, and the doped element unit cell model. For the supercell, grain boundary, and surface models doped with metallic elements, optimization was performed while keeping the lattice constants fixed. Using these computational parameters and models, the grain boundary energy for the yttrium surface [10$\bar{1}$0] was calculated to be 0.94 J/m², which is in good agreement with the value of 0.96 J/m² obtained by Tran et al.[39] using first-principles calculations.

Table 1   Crystal Structure Parameters at 0 K

| | | a/Å | c/Å | V/Å³ | Energy/eV | Source |
|---|---|---|---|---|---|---|
| hcp-Y | Contrast | 3.65543 | 5.64701 | 65.347 | -12.87 | This article |
| | | 3.6482 | 5.7318 | 66.066 | - | Ref [47] |
| | Deviation | 0.2% | 1.5% | 1.1% | - | - |

| | | $\gamma_{Sur}$(J/m²) | | | Source | |
|---|---|---|---|---|---|---|
| Surface [10$\bar{1}$0] | Contrast | 0.94 | | | This article | |
| | | 0.96 | | | Ref [39] | |
| | Deviation | 2.1% | | | - | |

| | x | y | z | $\gamma_{GB}$(J/m²) | $E_{WOS}$(J/m²) |
|---|---|---|---|---|---|
| GB{10$\bar{1}$0} | <$\bar{1}$2$\bar{1}$0> | <0001> | <10$\bar{1}$0> | 2.31 | 2.06 |

**Solution Energy**

The solution energy is defined as the energy required for a metallic element *X* to dissolve into bulk Y, and can be calculated using equation (1)[40].

$$E_{Sol}^{X} = E_{XY_{n-1}} - (n-1)E_Y - E_X \quad (1)$$

A positive solution energy indicates that the solute element $X$ requires energy to dissolve into bulk Y, whereas a negative solution energy suggests that energy is released, implying that the corresponding structure is stable. Here, $E_{XY_{n-1}}$ is the total energy of bulk Y containing the solute element X after full relaxation, while $E_X$ and $E_Y$ represent the energies of a single atom of the solute element $X$ and bulk Y, respectively. $n$ denotes the number of atoms in the bulk metal Y. Although the atomic radii of the solute elements studied are smaller than that of Y, these solute elements cannot occupy interstitial positions in bulk Y. Therefore, in the calculations related to doping in the bulk and subsequent grain boundary models, a substitutional doping method is employed, where the solute elements replace Y atoms.

**Segregation Energy**

The segregation energy is defined here as the energy required for a solute element $X$ to segregate from bulk Y to the vicinity of the grain boundary, as given by equation (2)[26,41].

$$E_{Seg}^X = E_{GB}^X - E_{GB} - (E_{bulk}^X - E_{bulk}) \quad (2)$$

Here, $E_{GB}^X$ and $E_{GB}$ denote the total energies after structural relaxation of the grain boundary with and without the metallic element $X$, respectively. Similarly, $E_{bulk}^X$ and $E_{bulk}$ represent the total energies after structural relaxation of bulk Y with and without the metallic element $X$. The segregation energy $E_{Sep}^X$ indicates the tendency of the metallic element to segregate from bulk Y to the vicinity of the grain boundary. A negative segregation energy implies that the solute atom prefers to segregate near the grain boundary, while a positive value suggests that the solute atom favors remaining in the bulk environment.

**Work Of Separation**

In Fig. 1, surface TOP and surface BOTTOM refer to the two surface models, FS1 and FS2, formed by the cleavage of the grain boundary. The cleavage path divides the grain boundary into two surfaces along the boundary line. The Work of Separation ($E_{WOS}^X$) represents the energy per unit area required to separate the grain boundary into two free surfaces, FS1 and FS2, as given by equations (3) and (4)[42,43].

$$E_{WOS}^X = \frac{E_{FS1} + E_{FS2}^X - E_{GB}^X}{S} \qquad (3)$$

$$E_{WOS} = \frac{E_{FS1} + E_{FS2} - E_{GB}}{S} \qquad (4)$$

A positive $E_{WOS}^X$ indicates that energy must be absorbed for cleavage to occur, whereas a negative value suggests that the grain boundary will spontaneously cleave toward a lower-energy surface. In the above equations, $E_{FS1}$ denotes the total energy of surface TOP, and $E_{FS2}^X$ and $E_{FS2}$ represent the energies of surface BOTTOM with and without solute element X, respectively, as referenced in equation (2). All energies mentioned have been fully relaxed through structural optimization.

**Strengthening Effect**

Based on the Rice-Wang model, the strengthening energy calculation is commonly used to characterize the grain boundary's resistance to tensile stress[44-46], as shown in equation (5).

$$E_{SE}^X = E_{GB}^X - E_{GB} - (E_{FS}^X - E_{FS}) \qquad (5)$$

A positive strengthening energy, indicated by larger values, signifies stronger embrittlement effects of the solute element on the grain boundary, leading to a reduction in grain boundary strength. Conversely, a negative or lower strengthening energy suggests that the solute element enhances the grain boundary's resistance to tensile stress, thereby improving its strength.

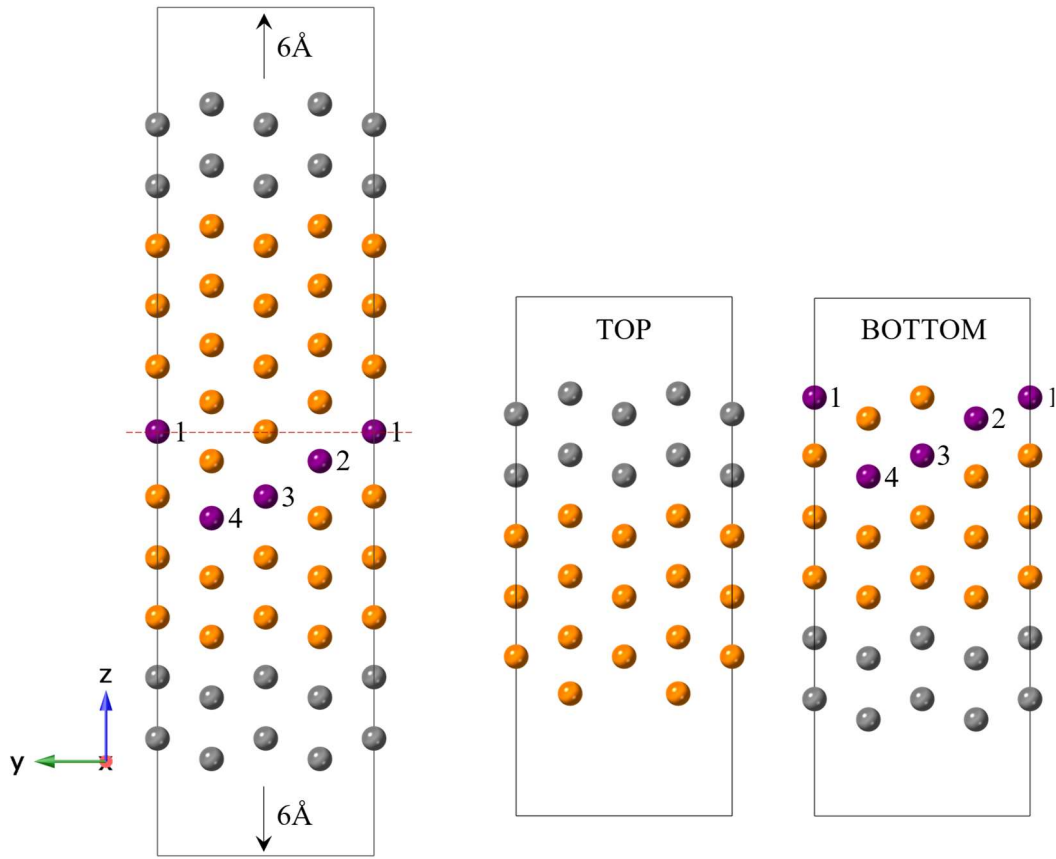

Fig.1　On the left, the grain boundary model of the Y twin grain boundary $\{10\bar{1}0\}$ is depicted. The red dashed line indicates the grain boundary line, with purple atoms representing different segregation sites of the solute atom, and gray atoms denoting fixed atoms. On the right, the upper and lower surfaces formed by cleavage along the grain boundary line are shown.

**Grain Boundary Energy**

Grain boundary energy serves as an indicator of the boundary's stability: the lower the grain boundary energy, the more stable the boundary. During material preparation and cooling, the formation of various grain boundaries is possible, and a lower grain boundary energy indicates a higher likelihood of the corresponding boundary appearing in the material. In this study, grain boundary energy is calculated according to equations (6) and (7), where S denotes the cross-sectional area of the grain boundary, and other symbols are consistent with those defined in previous equations.

$$\gamma_{GB}^X = \frac{E_{GB}^X - (n-1)E_{bulk} - E_X}{S} \quad (6)$$

$$\gamma_{GB} = \frac{E_{GB} - nE_{bulk}}{S} \quad (7)$$

**Results and discussions**

**Solution Effect**

The solution energies of 34 solute elements in bulk Y are depicted in Fig.2. Among the 3d, 4d, and 5d transition metals, with the exceptions of Rh, Pd, Cd, Ir, Pt, Au, and Hg, other elements require the absorption of a certain amount of energy to dissolve in bulk Y, indicating the instability of the corresponding structures. Conversely, for the seven main group elements, except for Be, the corresponding solution energies are all less than 0, indicating that these elements form stable structures when dissolved in bulk Y. To investigate the correlation between solution energy and atomic radius, the relative atomic radius is expressed as the absolute value of the difference between the atomic radius of the solute element and that of the Y element, i.e., $\Delta R = |R_X - R_Y|$, where $R_X$ and $R_Y$ represent the atomic radii of the solute element $X$ and Y element, respectively. The atomic radius of Y is 1.82 Å.

By observing Fig.2, it can be seen that higher solution energies often correspond to larger relative atomic radii, such as those observed for the 3d transition metals V and Mn, the 4d transition metals Mo and Te, and the main group metal Be. However, for the 4d transition metal Pd and the 5d transition metals Au, Pt, and Ir, despite having larger relative atomic radii compared to Y, their corresponding solution energies are less than 0. This suggests that the influence of the relative atomic radius of different solute elements on their solution energy in bulk Y is not entirely dominant. The properties of solute elements also play a crucial role. Typically, the electronegativity of an element is used to define its ability to attract electrons in a compound, with higher electronegativity values indicating a stronger ability to attract electrons. The relative electronegativity is expressed as $\Delta\chi = |\chi_X - \chi_Y|$, where $\chi_X$ and $\chi_Y$ represent the electronegativities of the solute element $X$ and the Y element, respectively (as shown in Fig.2). Further analysis of the correlation between the relative atomic radius, relative electronegativity of solute elements, and their corresponding solution energy reveals correlation coefficients of 0.57 and -0.22, respectively. This indicates a moderate positive correlation between the solution energy of solute element $X$ in bulk Y and the relative atomic radius, and a weakly negative correlation with relative electronegativity. When these correlations are refined to different groups, as shown in Table 2, main group elements exhibit a stronger correlation, while other group elements show a weaker correlation. This suggests that the factors influencing the solution of solute elements in bulk Y are not solely the relative atomic radius and relative electronegativity. Therefore, it is insufficient to determine the solution trends based solely on relative atomic radius and electronegativity.

In Fig.2, the solution energies of 4d and 5d transition metals exhibit a consistent trend among elements within the same group across different periods. For example, from Zr(Hf) to Mo(W), the solution energy increases, then decreases from Mo(W) to Pd(Pt), and increases again from Pd(Pt) to Ag(Au). However, deviations are observed for Cd and Hg, which may be related to their liquid metal properties. Starting with Rh(Ir), the solution energy becomes negative. In contrast, the 3d transition metals show different solution trends due to the magnetic properties of elements such as Cr, Mn, Fe, Co, and Ni.

Table 2  Correlation coefficients of the solution energy of solute elements in bulk Y with relative atomic radius and relative electronegativity.

|  | 3d | 4d | 5d | main |
| --- | --- | --- | --- | --- |
| $\Delta R$ | -0.53 | -0.17 | -0.24 | -0.90 |
| $\Delta\chi$ | 0.17 | -0.04 | -0.27 | -0.78 |

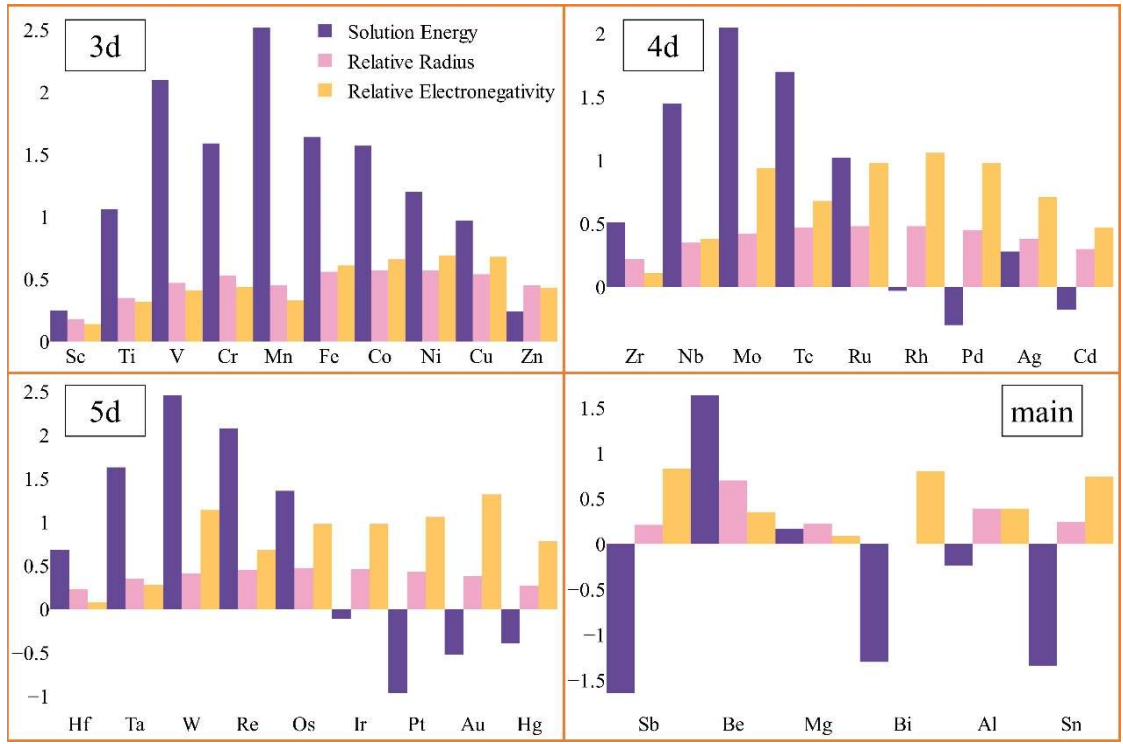

Fig.2　Solution energies of different solute elements in bulk Y. Purple indicates solution energy, and pink indicates a relative atomic radius. The groups are represented as 3d, 4d, 5d, and main.

**Segregation Effect**

According to equation (2), the segregation energies of 34 solute elements at different sites near the twin grain boundary(GB) $\{10\bar{1}0\}$ are shown in Fig.3. It is observed that the segregation abilities of all solute elements differ at various sites, but all metal elements can segregate to the vicinity of the GB. Except for Mn, the segregation energies of the other 33 solute elements at site-1 are all significantly greater than 0, with the smallest value being 0.19eV, indicating that these elements are not easily segregated to site-1. The segregation energies of all elements at site-2 are less than 0, indicating that solute elements are more likely to segregate at site-2 compared to site-1. At site-3, except for the main group elements Al, Bi, Mg, Sb, and Sn, which have segregation energies significantly greater than 0, the segregation energies of other elements are close to 0, with the 3d, 4d, and 5d transition metals exhibiting oscillations around 0. At site-4, the segregation energies of all transition metals are less than 0, indicating that these elements can segregate at this site. Main group elements such as Bi, Mg, Sb, and Sn are relatively difficult to segregate at site-4, similar to their behavior at site-2. Overall, when solute elements are doped near the grain boundary at sites 1-4, most elements show a sequential decrease in segregation energy at site-1, site-3, site-4, and site-2. This indicates that when solute elements segregate from bulk Y to site-4 near the $\{10\bar{1}0\}$ twin grain boundary, there is a tendency to further segregate to site-2 from an energy perspective, with site-3 acting as an energy barrier that obstructs segregation. Conversely, some elements exhibit a sequential decrease in segregation energy at site-1, site-3, site-2, and site-4, indicating that after segregating to site-4, the energy does not support segregation towards locations closer to the grain boundary. This pattern is observed for elements such as Fe among the 3d transition metals, Tc and Ru among the 4d transition metals, and Re, Os, and Ir among the 5d transition metals. In practice, dopant elements generally appear near the grain boundary line (site-1 and site-2) during the formation of Y-based alloys, and then segregate towards the grain boundary within the alloy. If elements are doped at site-1 of the grain boundary, they will segregate towards the site with lower segregation energy, which is site-2. Although elements such as Fe and Tc have the lowest segregation energy at site-4, site-3 still acts as an energy barrier, preventing diffusion to this site. If elements are doped at site-3 of the grain boundary, they will segregate towards site-4 or site-2. When elements segregate from the bulk to the grain boundary, site-3 acts as an energy barrier after segregation to site-4, preventing further segregation towards locations near the grain boundary. Therefore, site-4 is a potential segregation site for solute elements.

From the overall trend, the transition metals exhibit similar trends in segregation energy with increasing atomic number within the same group, and the interaction between the dopant atoms and substrate atoms varies depending on the atomic environment. For instance, 4d and 5d metals at site-1 show an initial increase followed by a decrease in segregation energy, at site-2 they show a decrease followed by an increase, at site-3 they oscillate around 0, and at site-4 they show a decrease followed by an increase. In contrast, the trends for 3d transition metals are disrupted by the magnetic properties of elements such as Cr, Mn, Fe, Co, and Ni.

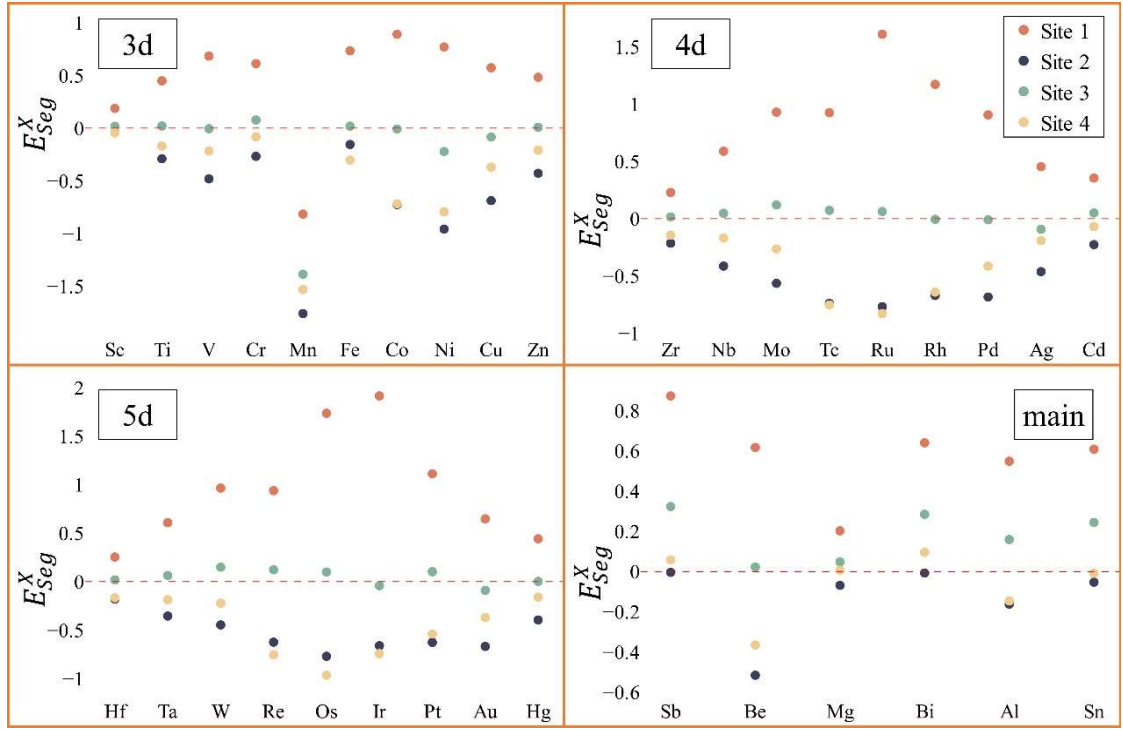

Fig. 3　The segregation energies of 34 solute elements at sites 1-4 near the $\{10\bar{1}0\}$

Among the transition metal elements, Mn demonstrates a significant segregation effect at the grain boundary, which strongly correlates with its limited solubility in bulk Y, as analyzed earlier. In contrast, the main group elements Bi, Sb, and Sn exhibit segregation energies close to zero at the most favorable segregation site (site-2), indicating a negligible tendency for segregation at the grain boundary. This observation is consistent with Fig. 2, where these elements display stable solution states in bulk Y. The solution energies of other solute elements are greater than those of the main group elements Bi, Sb, and Sn, corresponding to a more pronounced segregation trend at site-4 and site-2. An analysis of the correlations between relative atomic radius, relative electronegativity, and segregation energy reveals that the correlation coefficients at sites 1–4 are 0.31 and -0.58 and -0.25 and -0.56 for relative atomic radius, and 0.66, -0.29, 0.18, and -0.28 for relative electronegativity, respectively. There is no strong correlation between these two factors and the corresponding segregation energies. Additionally, the different atomic environments at various segregation sites lead to significant differences in the correlation coefficients at each site.

**Work Of Separation**

Based on equations (3) and (4), the value of the Y $\{10\bar{1}0\}$ twin grain boundary was calculated to be 2.06J/m², indicated by the red dashed line in Fig.4. The corresponding values for the segregation of 34 solute elements are also shown in Fig.4. It is evident that the segregation of these 34 solute elements at site-4 near the grain boundary increased $E_{WOS}^{X}$ (work of separation). The degree of increase was the smallest for Sc, at 0.01J/m², and the largest for Mn, at 0.37J/m². At segregation site-1, site-2, and site-3, the segregation of most solute elements decreased the value, such as the 3d transition metals and main group elements. However, the segregation of the 4d transition metal Zr at site-1 and site-2 near the grain boundary, the 3d and 5d transition metals Cr, Hf, and Ta at site-1, and Mn at site-3 increased the value.

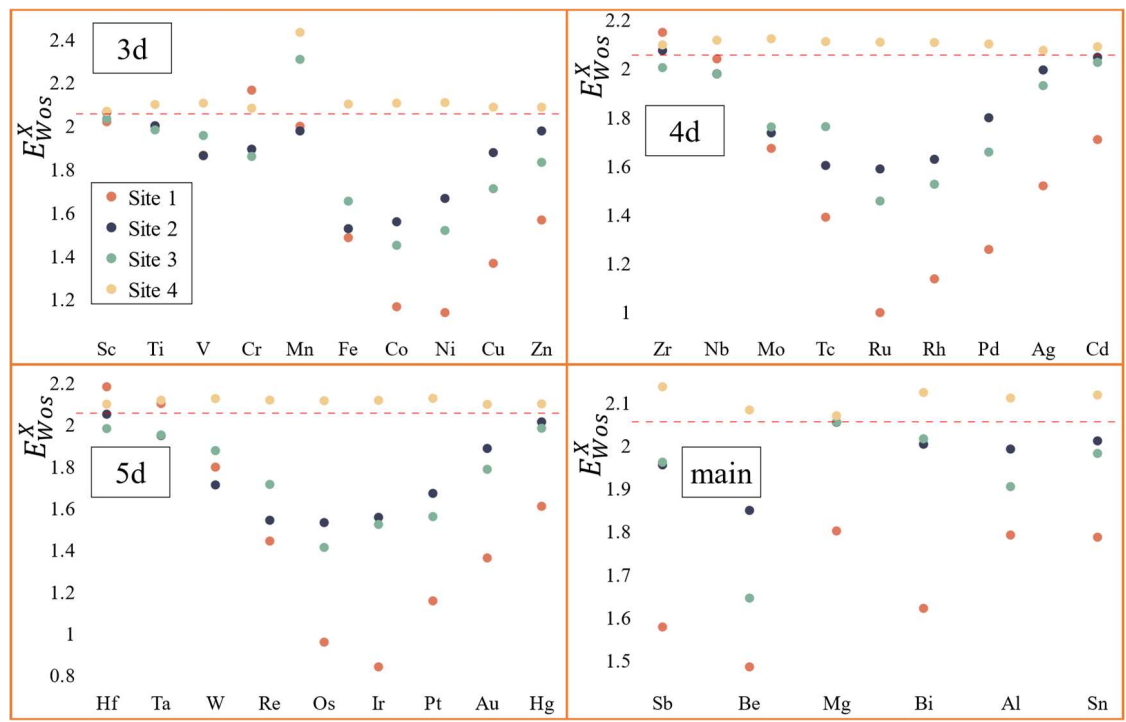

Fig.4  Work of separation corresponding to solute segregation at grain boundary sites 1–4. The red dashed line represents the work of separation for the pristine grain boundary.

**Strengthening Effect**

According to equation (5), the strengthening energies of thirty-four solute elements at grain boundary were calculated, as shown in Fig.5. It is clearly observed that after the segregation of all solute elements to site-4 near the grain boundary, their strengthening energies are all negative, indicating that they have a strengthening effect on the grain boundary, with Mn being the most significant. For sites 1-3, the main group solute elements exhibit an embrittlement effect on the grain boundary, with the exception of the 4d transition metal Zr, which strengthens the grain boundary at site-1 and site-2. Similarly, the 3d and 5d transition metals Cr, Hf, and Ta also strengthen the grain boundary at site-1, and among all solute elements, only Mn strengthens the grain boundary at site-3. The effect of these solute elements on grain boundary strengthening or embrittlement at different segregation sites corresponds to their fracture work relative to the pure grain boundary.

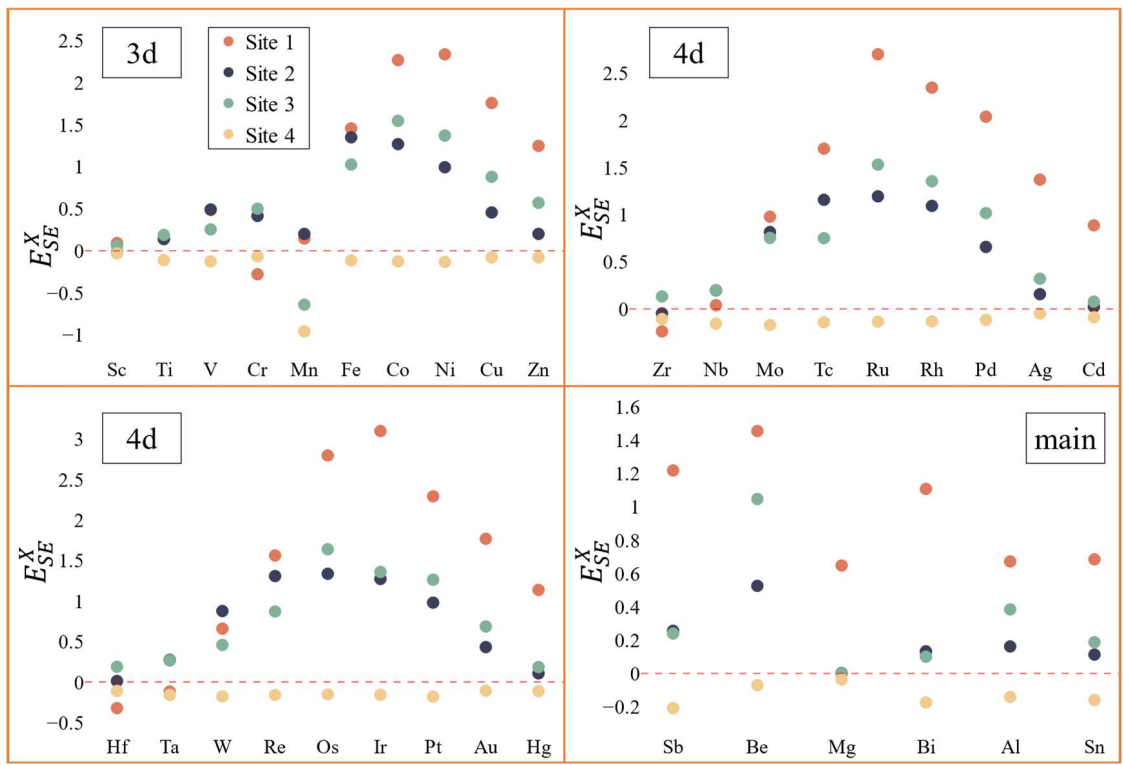

Figure 5: Strengthening energies of solute elements after segregation to four sites near the Y $\{10\bar{1}0\}$ twin boundary

**Segregation and Strengthening Effects**

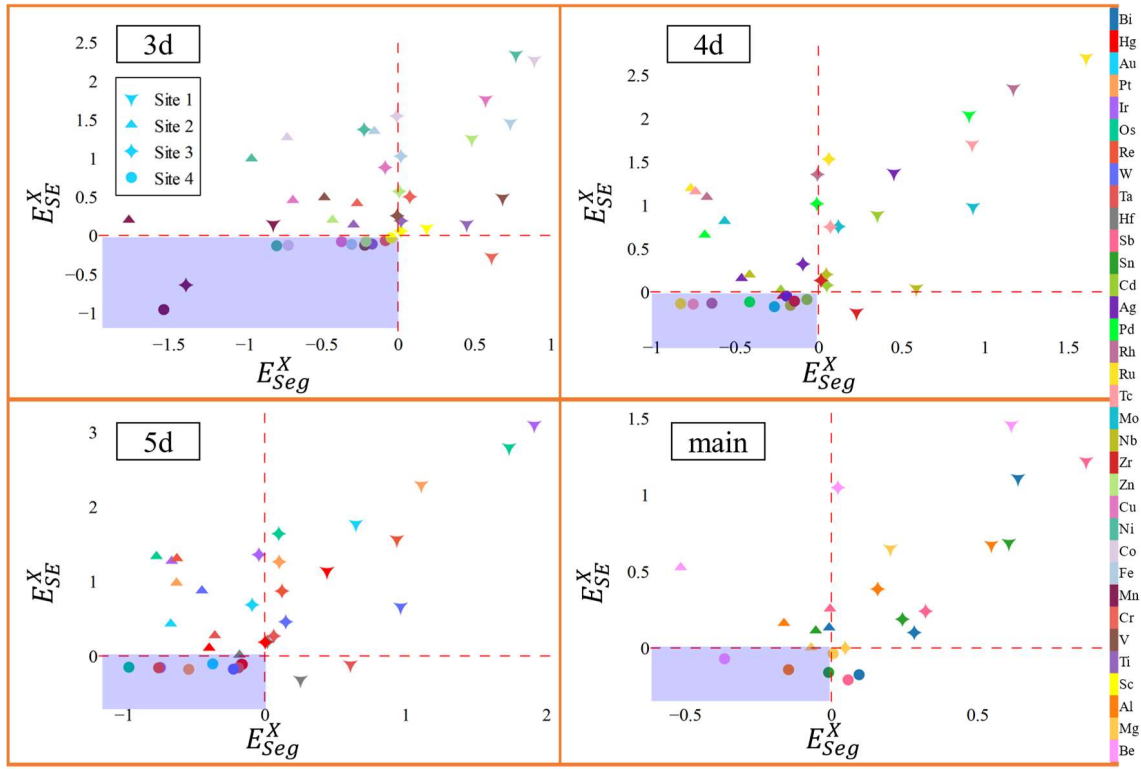

Fig.6 Schematic diagram of the segregation energies and strengthening energies of different solute elements (The four shapes in the legend represent corresponding segregation sites, and different colors represent different solute elements)

To gain a clearer understanding of the segregation trends of solute elements near the grain boundary and their effects on grain boundary strengthening or embrittlement after segregation, both aspects are illustrated in Fig.6. The horizontal and vertical axes represent segregation energy and strengthening energy, respectively. Different solute elements are indicated by corresponding colors, and the four sites near the grain boundary are marked with differently shaped symbols. It can be observed that Mn in the 3d transition metals easily segregates at site-3 and site-4 and strengthens the grain boundary after segregation, whereas Ni, Co, Cu, Fe, Zn, V, Ti, Cr, and Sc, although they easily segregate to site-4 near the grain boundary, do not show as significant a strengthening effect as Mn. Among the 4d transition metals, only Zr at site-2 shows both easy segregation and grain boundary strengthening properties. Other 4d transition metals and all 5d transition metals can segregate to site-4 near the grain boundary and strengthen the grain boundary but do not show a strengthening effect at other sites. In contrast, among the six main group elements, only Be and Al exhibit segregation and grain boundary strengthening effects, and only at site-4.

**Stability Of Grain Boundary**

Fig.7a describes the grain boundary energies of solute elements that can segregate near the grain boundary and strengthen it. The grain boundary energy was calculated based on equations (7) and (8), with the value of the pure Y$\{10\bar{1}0\}$ twin grain boundary being 2.31J/m², indicated by the red dashed line in the figure. The figure shows that only 11 solute elements can reduce the grain boundary energy of the original grain boundary, thereby enhancing its stability. These solute elements are Ag, Al, Au, Cd, Hg, Ir, Pd, Pt, Rh, Sn, and Zn, and their segregation sites are all at site-4 near the grain boundary. The element Pt has the strongest stabilizing effect on the grain boundary, with a grain boundary energy value of 1.67 J/m², while Ag has the smallest effect, with a grain boundary energy value of 2.29 J/m². The segregation of solute elements Sc, Zr, and Mn at site-2 and site-3 near the grain boundary increased the grain boundary energy of the original Y $\{10\bar{1}0\}$ twin grain boundary.

In the grain boundary energy curve for site-4 in Fig.7a, shaded regions mark the 3d, 4d, and 5d transition metal regions. For the 4d transition metals, we observe a trend in the influence on initial grain boundary energy as atomic number increases, characterized by a "rise→fall→rise" pattern. Specifically, as we move through the 4d transition series in order, the first three solute elements (Zr→Nb→Mo) increase the grain boundary energy after segregation, the next five elements (Mo→Tc→Ru→Rh→Pd) progressively decrease it, and the final two elements (Pd→Ag) increase it again. A similar trend appears for the 5d transition metals, where the first three solute elements (Hf→Ta→W) raise the grain boundary energy, the following five (W→Re→

Os→Ir→Pt) progressively lower it, and the last two (Pt→Au) increase it again. Examining the changes in outer-shell electron numbers, we find that in the 4d transition series, elements from groups IVB to VB to VIB show increasing unpaired electrons: from $4d^2$ to $5s^14d^4$ to $5s^14d^5$. Following this, in groups VIB to VIIB to VIIIB, the number of unpaired electrons decreases sequentially as follows: $5s^14d^5$ to $4d^5$ to $5s^14d^3$ to $5s^14d^2$ to zero. Subsequently, the unpaired electron count shifts as $0 \rightarrow 5s^1 \rightarrow 0$ in the VIII to IB to IIB groups. This pattern suggests that the increase in unpaired electrons correlates with higher grain boundary energy, whereas a decrease lowers it. Therefore, Cd in the 4d series shows a lower grain boundary energy than Ag due to its fewer unpaired electrons. Additionally, the 4d and 5d elements exhibit a similar pattern, as they belong to the same groups in the transition metals. Notably, the 4d element Cd and the 5d element Hg, both from the same group, differ in their trend of grain boundary energy reduction, possibly due to Hg's liquid metal characteristics. Conversely, the trend for the 3d transition metals (Cr, Mn, Fe, Co, and Ni) deviates from this pattern, although a similar overall tendency is observed. These elements are magnetic materials, which suggests that magnetism could influence their bonding with Y atoms. The observed effect of transition metals on grain boundary energy aligns with the solution trends discussed in the section on solubility effects.

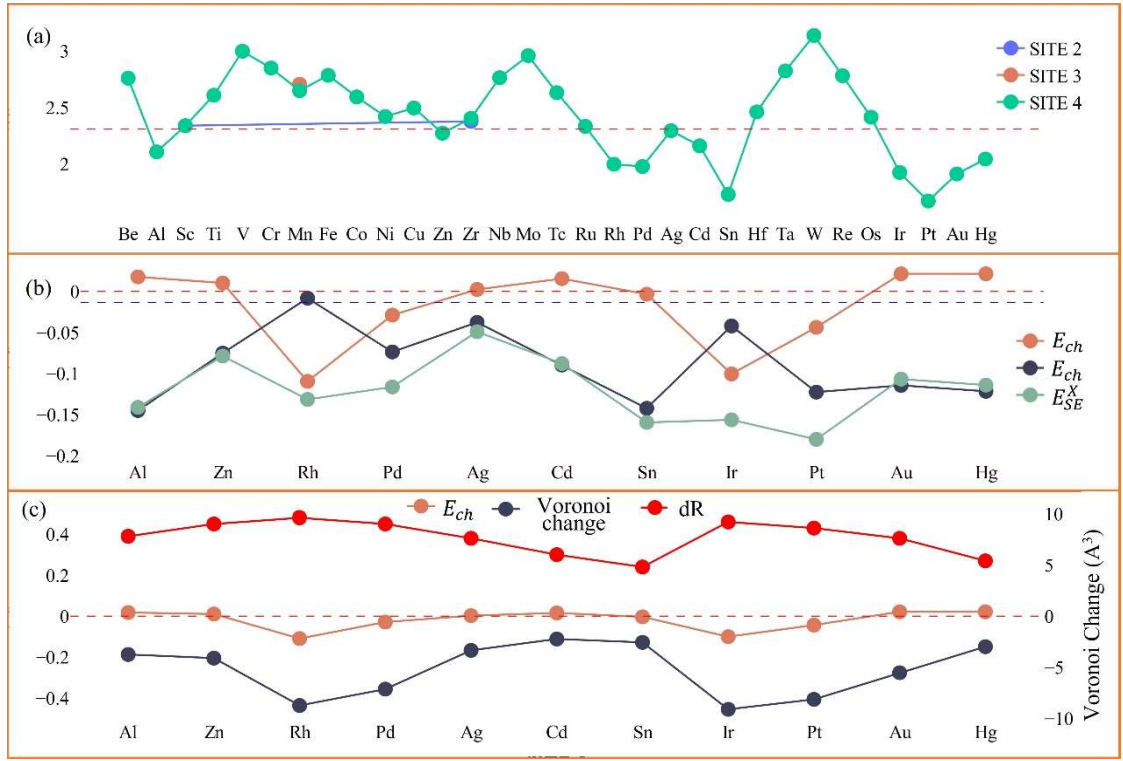

Figure 7: (a) Grain boundary energy chart for elements that can segregate to and strengthen the grain boundary; (b) Mechanical and chemical contributions to grain boundary strengthening for various elements; (c) Physical contributions, Voronoi volume change, and relative atomic radius (mechanical contribution energy and relative atomic radius correspond to the left y-axis, while Voronoi volume change corresponds to the right y-axis)

**Mechanical And Chemical Contributions**

The strengthening energy of the grain boundary after solute element segregation can be decomposed into three components: physical contribution energy, chemical contribution energy, and vacancy formation energy[18,22]. This section isolates each of these contributions to analyze their respective effects on grain boundary strengthening or embrittlement. In Fig.7, panels A and D represent the Y$\{10\bar{1}0\}$ twin grain boundary model with and without solute element segregation, respectively, along with the resulting free surfaces, Top and Bottom, after fracture. Panel B shows the grain boundary and Top and Bottom surfaces with one Y atom removed from the segregation site, based on panel A. Panel C is similar to panel B, depicting the grain boundary and surfaces after removing the solute element from panel D.

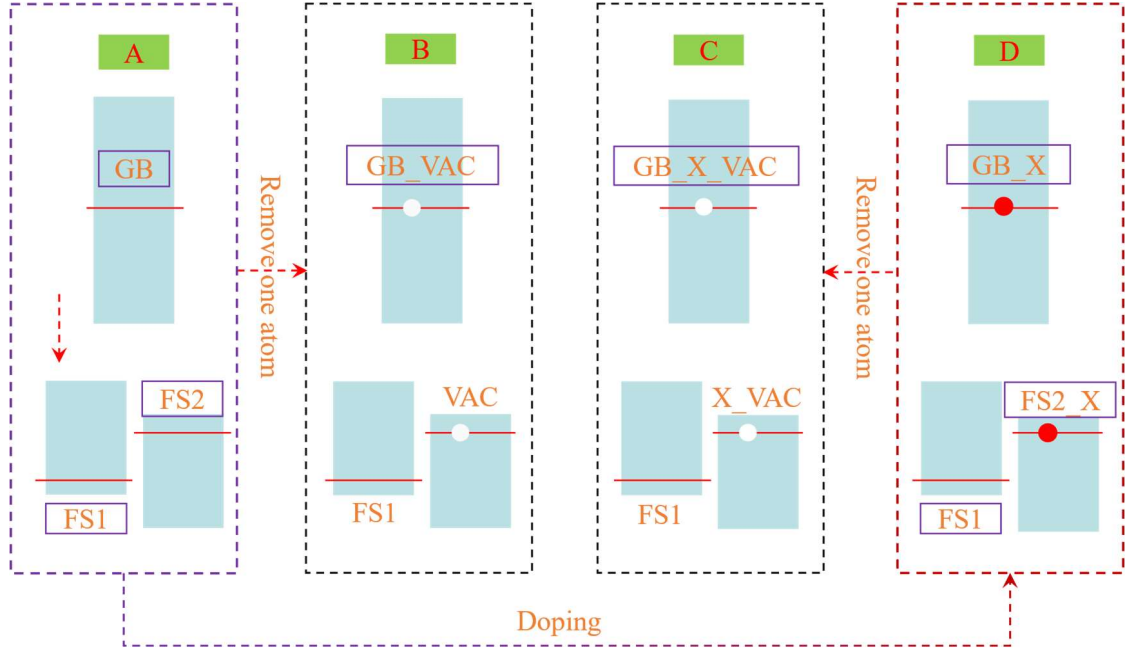

Fig.7　Schematic Diagram of Strengthening Energy Decomposition into Physical Contribution Energy, Chemical Contribution Energy, and Vacancy Formation Energy. (In each subfigure, the red solid line represents the grain boundary line; in subfigures B and C, white circles indicate vacancies after the removal of the respective atoms, and in subfigure D, red circles represent solute elements)

Based on Fig.7, the separation energy of the grain boundary($E_{Sep}$) is defined as the energy required to split the grain boundary into two free surfaces, $E_{Sep}^{label}$ represents the separation energy, the models in Fig.7 are labeled as A, B, C, and D, $E_{GB}$, $E_{BOTTOM}$, $E_{TOP}$ represent the energy of grain boundary $\{10\bar{1}0\}$ and the free surfaces Fs1 and Fs2, respectively, with B and C models reflecting static energies after atomic removal. The separation energy formula is provided in equation (8). Mechanical contribution energy, $E_{me}$, is defined as the energy contribution resulting from the re-equilibration of the Y atom environment after solute segregation, as shown in equation (9). Vacancy formation energy, $E_R$, represents the energy change after removing one Y atom from the pure grain boundary, following equation (10). Finally, the chemical contribution energy, $E_{ch}$, reflects the energy change due to solute segregation after accounting for physical and vacancy formation energies, as given by equation (11)[26,41,42].

$$E_{Sep}^{label} = E_{TOP} + E_{BOTTOM} - E_{GB} \qquad (8)$$

$$E_{me} = E_{Sep}^{B} - E_{Sep}^{C} \qquad (9)$$

$$E_{R} = E_{Sep}^{A} - E_{Sep}^{B} \qquad (10)$$

$$E_{ch} = E_{Sep} - E_{me} - E_{R} = E_{C} - E_{D} \qquad (11)$$

Based on the preceding section, the solute elements identified as capable of segregating to the grain boundary, strengthening the boundary, and enhancing its stability include two main-group elements (Al and Sn) and nine transition metals: Zn, Rh, Pd, Ag, Cd, Ir, Pt, Au, and Hg. Fig.7b shows the strengthening energy of these elements and their mechanical and chemical contribution components. In this figure, mechanical contribution energy, chemical contribution energy, and strengthening energy are represented by orange-red, black, and green lines, respectively. A positive value for mechanical or chemical contribution energy indicates that the solute element's segregation embrittles the original boundary strength, while a negative value indicates boundary strengthening. The blue dashed line represents the vacancy formation energy, which is -0.01 eV. Its embrittling effect on the grain boundary strength is minimal and thus is neglected here. By examining the figure, it can be observed that the chemical contribution energy for all metal elements is less than zero, indicating that the chemical contribution energy of these elements plays a strengthening role at the grain boundary. Rh has the smallest strengthening effect among all the solute elements, with a chemical contribution energy of -0.008 eV, while Al provides the most substantial strengthening effect, with a chemical contribution energy of -0.145 eV. For transition metals in the same group, such as Rh and Ir, the mechanical contribution energy primarily contributes to strengthening the grain boundary. For other elements, however, the chemical contribution energy is the main component in strengthening the grain boundary, which explains why the trend of the strengthening energy curve closely follows that of the chemical contribution energy curve.

Voronoi volume refers to the geometric region around a given point in an atomic coordinate system, where the surrounding space is enclosed by polyhedra formed by areas closest to that point. When dopant atoms are introduced into the grain boundary, lattice mismatch occurs, causing atoms to seek the most stable positions and adjust the surrounding space to release stress, causing a change in Voronoi volume. As shown in Fig.7c, the orange-red, black, and red curves represent the mechanical contribution energy, Voronoi volume change, and relative atomic radius, respectively. It can be observed that all three display consistent trends.

**Density of states**

The density of states refers to the number of available electronic states within a given energy range. It is an important characteristic of a material's electronic structure and is commonly used to describe the distribution of electronic states within a specific energy range. The density of states between the solute atoms and their nearest neighbor Y atoms at the Y$\{10\bar{1}0\}$ twin grain boundary after the segregation of 11 solute elements was calculated, as shown in Fig.8. By observing Fig.8, it can be seen that for elements like Ir, Rh, and Pt, the overlap between their d-orbitals and the d-orbitals of the nearest neighbor Y atoms is quite significant, indicating a large degree of hybridization between the d-orbitals. Considering their electron configurations, Ir is [Xe]$4f^{14}5d^{7}6s^{2}$, Rh is [Kr]$4d^{8}5s^{1}$, and Pt is [Xe]$4f^{14}5d^{7}6s^{2}$. It is apparent that these three solute elements have incomplete d-orbitals, which means that there is space in these orbitals to accept electrons or participate in bonding. This characteristic enhances the charge transfer and bonding capability between the solute atoms and the surrounding Y atoms, and the unfilled d-orbitals enable these elements to engage in charge transfer or sharing, thereby adjusting the charge density and electronic chemical potential at the grain boundary, which further reduces the grain boundary energy.

In contrast, Ag, Au, Cd, Hg, and Zn exhibit different hybridization degrees. Since the d orbitals of these elements are fully occupied by electrons, they primarily hybridize with the d orbitals of Y through weaker s and p orbitals, contributing limitedly to the density of states (DOS). Although these metals increase the DOS near the Fermi level, their overall effect in reducing grain boundary energy is weaker than that of metals with partially filled d orbitals. The d orbital states are mainly distributed in the energy region less than -3eV, with a significant distance from the Fermi level, lacking sufficient electrons to participate in hybridization. Thus, these elements exhibit relatively ineffective mechanisms in lowering the grain boundary energy. Additionally, the DOS diagrams reveal that elements within the same group of 4d and 5d transition metals show similar orbital interactions. The p orbitals of the main group metals Al and Sn display high electron density near the Fermi level, indicating that these electrons play an important role in conductivity. The strong hybridization between the p orbitals of Al and Sn and the d orbitals of Y, especially near the Fermi level (centered at 0eV), significantly alters the electron distribution, possibly forming covalent bonds, further lowering the grain boundary energy.

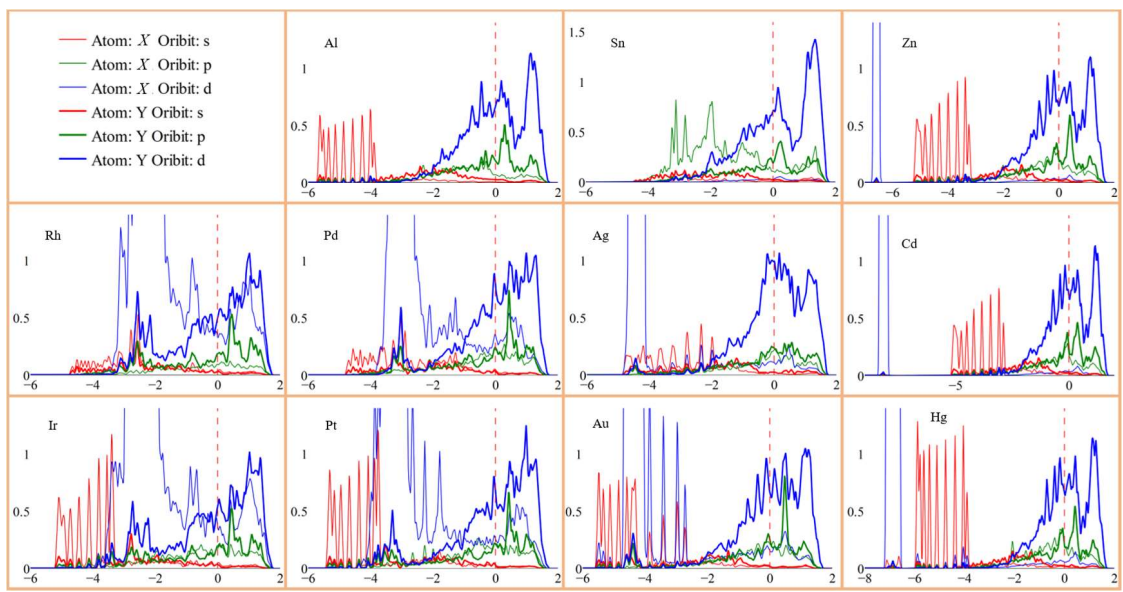

Fig. 8  Density of states (DOS) near the Fermi level for segregated elements and their nearest neighbor Y atoms

**Summary**

This study employed first-principles calculations to investigate the segregation and strengthening effects of 34 metallic elements near the grain boundary of rare earth metal Y. It was found that all solutes can segregate to the vicinity of the grain boundary. The segregation energies at site 3 indicate that this site can act as an energy barrier, preventing solute elements from segregating closer to the grain boundary from the bulk. According to the Rice-Wang model, the strengthening energy results indicate that all elements can strengthen the grain boundary. The 4d and 5d transition metals exhibit similar trends, showing consistent changes in solution trends and grain boundary energies within the same group of elements across different periods. Combining segregation energy, strengthening energy, and grain boundary energy, it was found that 11 elements (Al, Zn, Rh, Pd, Ag, Cd, Sn, Ir, Pt, Au, Hg) can segregate near the grain boundary, strengthen it, and enhance its stability. Additionally, by decomposing the strengthening energy into mechanical contribution energy, chemical contribution energy, and vacancy formation energy, it was found that the chemical contribution energy plays a major role in strengthening. The mechanical contribution energy of transition metal elements is related to the Voronoi volume change and relative atomic radius change of solute elements. Density of states analysis revealed that the stability enhancement of grain boundary by solute elements is due to orbital hybridization, forming more stable electronic states. Overall, this study provides a comprehensive perspective on the complex factors influencing the strength of hcp-Y grain boundary, making a significant contribution to the fields of materials science and metallurgical engineering.

**Acknowledgments**

A part of this research work is supported by University of Science and Technology Beijing High performance Computing university-level public platform.

# References

[1] Wang G., Zuo L., Esling C., Computer Simulation on the Tendency of Intergranular Fracture in Textured Polycrystalline Materials[J]. Philosophical Magazine A: Physics of Condensed Matter, Structure, Defects and Mechanical Properties, 2002, 82(12): 2499-2510.

[2] KHALAJHEDAYATI A, PAN Z, RUPERT TJ.Manipulating the Interfacial Structure of Nanomaterials to Achieve a Unique Combination of Strength and Ductility[J].Nature Communications,2016, 7.

[3] KING A, JOHNSON G, ENGELBERG D, et al.Observations of Intergranular Stress Corrosion Cracking in a Grain-Mapped Polycrystal[J].Science,2008, 321 (5887): 382.

[4] D. Raabe, M. Herbig, S. Sandlöbes, et al. Grain boundary segregation engineering in metallic alloys: A pathway to the design of interfaces,Current Opinion in Solid State and Materials Science, 2014, 18, 4, 253-261

[5] RAABE D, SANDLöBES S, MILLáN J, et al.Segregation Engineering Enables Nanoscale Martensite to Austenite Phase Transformation at Grain Boundaries: A Pathway to Ductile Martensite[J].Acta Materialia,2013, 61 (16): 6132-6152.

[6] D. Raabe, S. Sandlöbes, J. Millán, et al, Segregation engineering enables nanoscale martensite to austenite phase transformation at grain boundaries: A pathway to ductile martensite,Acta Materialia,Volume 61, Issue 16,2013,6132-6152,

[7] G. Da Rosa, P. Maugis, A. Portavoce, et al, Grain-boundary segregation of boron in high-strength steel studied by nano-SIMS and atom probe tomography, Acta Materialia, Volume 182, 2020, Pages 226-234

[8] Zhao D, Løvvik O M, Marthinsen K, et al. Segregation of Mg, Cu and their effects

on the strength of Al Σ5(210)[001] symmetrical tilt grain boundary[J]. Acta Materialia, 2018, 145: 235-246.

[9] Xiao Z, Hu J, Zhang X, et al. Uncovering the Zr segregation behavior and its effect on the fracture strength of Al Σ5(210)[100] symmetrical tilt grain boundary: Insight from first principles calculation[J]. Materials Today Communications, 2020, 25: 101268.

[10] Lu G-H, Zhang Y, Deng S, et al. Origin of intergranular embrittlement of Al alloys induced by Na and Ca segregation: Grain boundary weakening[J]. Physical Review B, 2006, 73 (22): 224115.

[11] Zhang S, Kontsevoi O Y, Freeman A J, et al. First principles investigation of zinc-induced embrittlement in an aluminum grain boundary[J]. Acta Materialia, 2011, 59 (15): 6155-6167.

[12] Xiao Z, Hu J, Liu Y, et al. Co-segregation behavior of Sc and Zr solutes and their effect on the Al Σ5(210)[110] symmetrical tilt grain boundary: a first-principles study[J]. Physical Chemistry Chemical Physics, 2019, 21 (35): 19437-19446.

[13] Xiao Z, Hu J, Liu Y, et al. Segregation of Sc and its effects on the strength of Al Σ5(210)[100] symmetrical tilt grain boundary[J]. Materials Science and Engineering: A, 2019, 756: 389-395.

[14] Razumovskiy V I, Lozovoi A Y, Razumovskii I M. First-principles-aided design of a new Ni-base superalloy: Influence of transition metal alloying elements on grain boundary and bulk cohesion[J]. Acta Materialia, 2015, 82: 369-377.

[15] Bentria E T, Lefkaier I K, Benghia A, et al. Toward a better understanding of the enhancing/embrittling effects of impurities in Nickel grain boundaries[J]. Scientific Reports, 2019, 9 (1): 14024.

[16] Bentria E L T, Lefkaier I K, Bentria B. The effect of vanadium impurity on

Nickel Σ5(012) grain boundary[J]. Materials Science and Engineering: A, 2013, 577: 197-201.

[17] Wang S, Xiong J, Zeng Q, et al. Effect of Nb on He segregation behavior in Ni Σ5 grain boundary: First-principles study[J]. Fusion Engineering and Design, 2020, 154: 111549.

[18] Liu W, Han H, Ren C, et al. Effects of rare-earth on the cohesion of Ni Σ5(012) grain boundary from first-principles calculations[J]. Computational Materials Science, 2015, 96: 374-378.

[19] Darling K A, Tschopp M A, VanLeeuwen B K, et al. Mitigating grain growth in binary nanocrystalline alloys through solute selection based on thermodynamic stability maps[J]. Computational Materials Science, 2014, 84: 255-266.

[20] Yang Y, Ding J, Zhang P, et al. The effect of Cr on He segregation and diffusion at Σ3(112) grain boundary in α-Fe[J]. Nuclear Instruments and Methods in Physics Research Section B: Beam Interactions with Materials and Atoms, 2019, 456: 7-11.

[21] Hu Y-J, Wang Y, Wang W Y, et al. Solute effects on the Σ3 111[1$\bar{1}$0] tilt grain boundary inBCC Fe: Grain boundary segregation, stability, and embrittlement[J]. Computational Materials Science, 2020, 171: 109271.

[22] Wang F. Y., Wu H. H., Zhou X. Y. et al, First-principle study on the segregation and strengthening behavior of solute elements at grain boundary in BCC iron. Journal of Materials Science & Technology, 2024, 189, 247-261

[23] Pei Z, Li R, Nie J-F, et al. First-principles study of the solute segregation in twin boundaries in Mg and possible descriptors for mechanical properties[J]. Materials & Design, 2019, 165, 107574.

[24] Hong Ju, Hong Ning, Zhao-Yuan Meng, et al. First-principles study on the

segregation behavior of solute atoms at {10$\bar{1}$2} and {10$\bar{1}$1} twin boundaries of Mg, Journal of Materials Research and Technology, 2023, 24, 8558-8571,

[25] Wu X., You Y. W., Kong X. S., et al. First-principles determination of grain boundary strengthening in tungsten: Dependence on grain boundary structure and metallic radius of solute[J]. Acta Materialia, 2016, 120, 315-326.

[26] Huang Z, Chen F, Shen Q, et al. Combined effects of nonmetallic impurities and planned metallic dopants on grain boundary energy and strength[J]. Acta Materialia, 2019, 166, 113-125.

[27] Paul C. Millett, R. Panneer Selvam, Ashok Saxena,Stabilizing nanocrystalline materials with dopants, Acta Materialia, 2007, 55, 7,

[28] Ito K., Sawada H., Ogata S.First-principles study on the grain boundary embrittlement of bcc-Fe by Mn segregation, Physical Review Materials, 2019, 3, 1

[29] Liang L, Hardouin Duparc O B M. First-principles study of four mechanical twins and their deformation along the c-axis in pure α-titanium and in titanium in presence of oxygen and hydrogen[J]. Acta Materialia, 2016, 110: 258-267.

[30] Zheng H., Li X., Tran R., et al. Grain boundary properties of elemental metals. Acta Materialia, 2019, 186, 40-49. doi:10.1016/j.actamat.2019.12.030

[31] Kresse G., Hafner J., Ab initio molecular dynamics for liquid metals[J]. Physical Review B, 1993, 47(1), 558-561. doi:10.1103/PhysRevB.47.558

[32] Kresse G., Hafner J., Ab initio molecular-dynamics simulation of the liquid-metal-eamorphoussemiconductor transition in germanium[J]. Physical Review B, 1994, 49(20), 14251-14269. doi:10.1103/PhysRevB.49.14251

[33] Kresse G., Furthmüller J., Efficient iterative schemes for ab initio total-energy

calculations using a plane-wave basis set[J]. Physical Review B, 1996, 54(16), 11169-11186. doi:10.1103/PhysRevB.54.11169

[34] Ask H. L. , Mortensen J. J., Blomqvist J., et al. The atomic simulation environment-a Python library for working with atoms[J]. Journal of physics. Condensed matter: an Institute of Physics journal, 2017, 29(27), 273002. doi:10.1088/1361-648X/aa680e

[35] Momma K., & Izumi, F., VESTA: a three-dimensional visualization system for electronic and structural analysis. J. Appl. Cryst. 2008, 41, 653-658. doi:10.1107/S0021889808012016

[36] Perdew J. P., Burke K. Ernzerhof M., Generalized Gradient Approximation Made Simple[J], Phys. Rev. Lett., 1996, 77(18), 3865-3868. doi: 10.1103/PhysRevLett.77.3865

[37] Kresse G., Furthmüller J., Efficient iterative schemes for ab initio total-energy calculations using a plane-wave basis set, Phys. Rev. B, 1996, 54, 11169. doi:10.1103/PhysRevB.54.11169

[38] Huang Z. F., Chen F., Shen Q., et al. Uncovering the influence of common nonmetallic impurities on the stability and strength of a Σ5(310) grain boundary in Cu[J]. Acta Materialia, 2018, 148, 110-122. doi:doi.org/10.1016/j.actamat.2018.01.058

[39] Tran R., Xu Z., Radhakrishnan B. et al. Surface energies of elemental crystals. Sci Data 3, 2016, 160080. doi:10.1038/sdata.2016.80

[40] Xue Z., Zhang X. Y., Qin J. Q., et al. Exploring the effects of solute segregation on the strength of Zr $\{10\bar{1}1\}$ grain boundary: A first-principles study, Journal of Alloys and Compounds, 2020, 812, 152153. doi: 10.1016/j.jallcom.2019.152153

[41] Huang Z. F., Wang P., Chen F. et al. Understanding solute effect on grain

boundary strength based on atomic size and electronic interaction[J]. Sci Rep 10, 2020, 16856. doi:10.1038/s41598-020-74065-1

[42] Shi Y., Xue H., Tang F., et al. Effects of transition metal segregation on the thermodynamic stability and strength of Ni Σ11 [110](113) symmetrical tilt grain boundary[J]. Vacuum, 2023, 212, 112036. doi:10.1016/j.vacuum.2023.112036

[43] Mai H.L., Cui X.Y., Scheiber D., et al. The segregation of transition metals to iron grain boundaries and their effects on cohesion[J]. Acta materialia, 2022, 231, 117902. doi:0.1016/j.actamat.2022.117902

[44] Wang J. B., Han H,, Tian Y. Z., et al. Exploring the influence of cosegregation on mechanical properties of Mg{$10\bar{1}1$} twin boundaries: A first-principles investigation[J]. Materials Today Communications, 2023, 37, 107520. doi:10.1016/j.mtcomm.2023.107520

[45] Rice J. R., Wang J. S., Embrittlement of interfaces by solute segregation. Materials Science and Engineering: A, 1989, 107, 23-40. doi:10.1016/0921-5093(89)90372-9

[46] Hui J., Zhang X., Liu T., et al. First-principles calculation of twin boundary energy and strength/embrittlement in hexagonal close-packed titanium[J]. Materials & Design, 2022, 213, 110331. doi:10.1016/j.matdes.2021.110331

[47] Spedding F. H., Hanak J. J., & Daane A. H., High temperature allotropy and thermal expansion of the rare-earth metals. Journal of the Less Common Metals, 1961, 3(2), 110-124. doi:10.1016/0022-5088(61)90003-0